\documentclass[aps,twocolumn,prb,showpacs]{revtex4-1}
\usepackage{graphics,bm}
\usepackage{amssymb,amsmath,amsfonts}
\usepackage{epsfig}
\usepackage{epsf}
\usepackage[breaklinks]{hyperref}

\def\be{\begin{equation}} \def\ee{\end{equation}}
\def\bea{\begin{eqnarray}} \def\eea{\end{eqnarray}}
\def\nn{\nonumber}
\def\Ham{\mathcal{H}}

\begin{document}

\title{Vacancy Driven Orbital and Magnetic Order in (K,Tl,Cs)$_y$Fe$_{2-x}$Se$_2$}

\author{Weicheng Lv}

\author{Wei-Cheng Lee}

\author{Philip Phillips}
\affiliation{Department of Physics, University of Illinois, 1110 West Green Street, Urbana, Illinois 61801, USA}

\date{\today}

\begin{abstract}
We investigate the effects of the $\sqrt{5}\times\sqrt{5}$ Fe vacancy
ordering on the orbital and magnetic order in
(K,Tl,Cs)$_y$Fe$_{2-x}$Se$_2$ using a three-orbital ($t_{2g}$)
tight-binding Hamiltonian with generalized Hubbard interactions.
We find that vacancy order enhances electron correlations, resulting in the onset of a block antiferromagnetic phase with large moments at smaller interaction strengths. In addition, vacancy ordering modulates the kinetic energy differently for the three $t_{2g}$ orbitals.  This results in a breaking of the degeneracy between the $d_{xz}$ and $d_{yz}$ orbitals on each Fe site, and the onset of orbital order.
Consequently, we obtain a novel inverse
relation between orbital polarization and the magnetic moment.
We predict that a transition from high-spin to low-spin states accompanied by a
crossover from orbitally-disordered to orbitally-ordered states
will be driven by doping the parent compound with electrons, which can be
verified by neutron scattering and soft X-ray measurements.
\end{abstract}
\pacs{74.70.Xa, 75.25.Dk, 71.27.+a, 71.30.+h}

\maketitle

\section{Introduction}
\begin{figure}
  \centering
  \includegraphics[width=8cm]{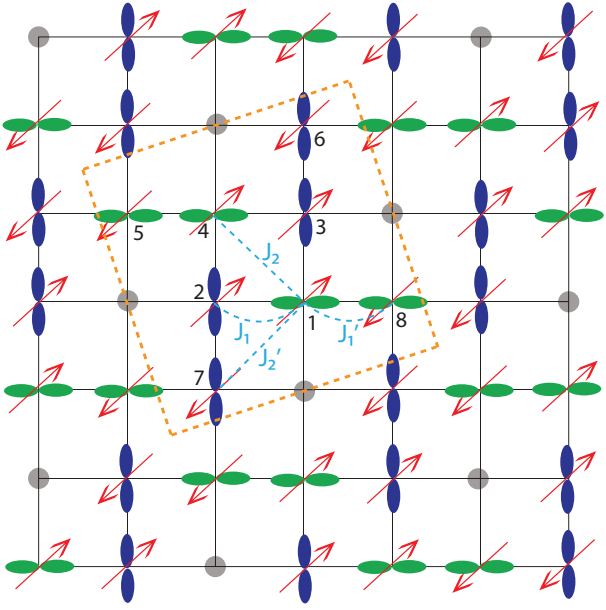}
  \caption{(Color online) Illustration of the $\sqrt{5}\times\sqrt{5}$ Fe vacancy order, subsequent block antiferromagnetic order, and possible orbital order due to the presence of vacancies. The Fe vacancies are represented by the gray circles. The dashed lines depict the unit cell that contains 8 Fe atoms and 2 Fe vacancies. $J_1$, $J_2$, $J_1^\prime$, and $J_2^\prime$ denote the intra- and inter-block nearest-neighbor and next-nearest-neighbor superexchanges, respectively.}
  \label{fig:schematic}
\end{figure}

At ambient pressure, FeSe undergoes a superconducting transition at
8K \cite{hsu_fc2008}.  However, when it is exposed to potassium, thallium or cesium, the
superconducting transition temperature roughly quadruples reaching a
value just above 30K \cite{guo_jg2010,mizuguchi2011,wang_af2011}.  Such a
significant increase in $T_c$ is due entirely to the effect K, Tl or Cs
have when they are intercalated between the FeSe layers, the primary
focus of this paper.  The average atomic ratios of K:Fe:Se are
0.39:0.85:1 \cite{guo_jg2010} in these new iron-chalcogenide
superconductors.  Hence, the new iron-based superconductors are iron
deficient.  In fact, the new wrinkle the
A$_y$Fe$_{2-x}$Se$_2$ (A=K, Tl, Cs) superconductors have introduced into the
field of novel superconductors is iron vacancy order.
The Fe vacancies form a $\sqrt{5} \times \sqrt{5}$ pattern
\cite{fang_mh2011,bacsa2011,zavalij2011}.  Such vacancy ordering occurs at a
higher temperature than the transition to the block antiferromagnetic
phase with an unusually large magnetic moment of 3.3$\mu_B$ per Fe
ion \cite{bao_w2011} (See Fig.~\ref{fig:schematic}). By contrast, FeSe is non-magnetic while FeTe$_{1-x}$Se$_x$ has moment up to 2.0$\mu_B$ in the non-superconducting state.  With such a large moment, one might
anticipate that the correlations in the new family of iron
chalcogenide superconductors are strongly enhanced.  In fact, they
are.
In sharp contrast to other families of iron-based superconductors,
A$_y$Fe$_{2-x}$Se$_2$ for $x>0.5$ are insulating \cite{fang_mh2011},
possibly of the Mott type.
This raises the possibility that superconductivity in these materials
is enhanced as a result of the increased number of unpaired
d-electrons that form the Mott insulating state, as has been proposed
recently \cite{wclee}.
Consistent with this picture is the experimental finding that
superconductivity is strongly suppressed with a small amount of Co
doping \cite{zhou_tt2011}.

Due to enhanced correlations in the A$_y$Fe$_{2-x}$Se$_2$
superconductors, local 3$d$ models with Hubbard-type interactions are required to describe the system \cite{yu_r2011,zhou_y2011}.
Consequently a vacancy-modulated $J_1$-$J_2$ model has been proposed \cite{cao2011} and applied to analyze the magnetic phase diagram \cite{lu_f2011,yu_r2011b,xu_b2011} for the insulating parent compounds. However, very few studies \cite{chen2011} consider both hopping and on-site interactions.
In this paper, we construct such a model exploiting a three-orbital ($t_{2g}$) tight-binding Hamiltonian with full
on-site Hubbard interactions. The vacancies are introduced with a $\sqrt{5}\times\sqrt{5}$ order which is observed in crystal X-ray diffraction studies
for K$_{0.8+y}$Fe$_{1.6-x}$Se$_2$ \cite{fang_mh2011,bacsa2011,zavalij2011},
corresponding to $20\%$ of the Fe vacancies.
We find that the vacancy ordering affects both the orbital and magnetic properties.
On the one hand, the presence of the vacancy lattice explicitly breaks the degeneracy between the $d_{xz}$ and $d_{yz}$ orbitals at  neighboring sites. For example, at site 1 of Fig.~\ref{fig:schematic}, the vacancy along the $-y$ direction leads to different modulations of the hopping terms for $d_{xz}$ and $d_{yz}$, due to their spatial anisotropy.
Without loss of generality, we will assume that the $d_{xz}$ orbital is favored at site 1. We can repeat this analysis for each Fe site, resulting in the particular orbital order as illustrated in Fig.~\ref{fig:schematic}, without turning on any interactions. This particular orbital order directly results in the breaking of local C$_4$ symmetry,
which has been hinted at by a recent NMR study \cite{torchetti2011}.
On the other hand, vacancy ordering also enhances the correlation effects by reducing the kinetic energy, leading to a block antimagnetic order at smaller interaction strengths.
These two tendencies compete with one  another as the system is doped
away from the insulating parent state, giving rise to a novel inverse
relation between the orbital polarization and the magnetic moment.

\section{Model}
We start from a three-orbital ($t_{2g}$) tight-binding model with the hopping parameters adopted from Ref.~\onlinecite{daghofer2010}.
The vacancies with a $\sqrt{5}\times\sqrt{5}$ order lead to an enlarged unit cell (see Fig.~\ref{fig:schematic}). The kinetic energy then takes the form
\bea
    \Ham_K = \sum_{\bm{k},i,j,\alpha,\beta,\mu} \xi_{ij}^{\alpha\beta}(\bm{k}) c_{i\alpha\mu}^\dagger(\bm{k}) c_{j\beta\mu}(\bm{k}),
\eea
where $c_{i\alpha\mu}^\dagger$ creates an electron on orbital $\alpha$ with spin $\mu$ at site $i$. We have $i=1,2,\dots,8$, as labeled in Fig.~\ref{fig:schematic}, and $\xi_{ij}^{\alpha\beta}(\bm{k}) = \sum t_{ij}^{\alpha\beta} \exp{[i\bm{k}\cdot(\bm{r}_j-\bm{r}_i)]}$, with $t_{ij}^{\alpha\beta}$ being the hopping amplitudes and $\bm{k}$ defined within the Brillouin zone of the enlarged unit cell.
We further consider the following Hubbard interactions on each site
\bea
\Ham_I &=& \sum_{\alpha} U \hat{n}_{\alpha\uparrow}\hat{n}_{\alpha\downarrow} + \sum_{\beta>\alpha}(V-\frac{J}{2}) \hat{n}_{\alpha} \hat{n}_{\beta}
- \sum_{\beta>\alpha}2J\vec{S}_{\alpha}\cdot \vec{S}_{\beta}\nn\\ &+& \sum_{\beta > \alpha}
J^\prime \big(c_{\alpha\uparrow}^\dagger c_{\alpha\downarrow}^\dagger c_{\beta\downarrow} c_{\beta\uparrow}+ h.c.\big),
\label{hamI}
\eea
where $U$, $V$, $J$ and $J^\prime$ are the intra- and inter-orbital Coulomb repulsion, Hund's coupling, and pair hopping, respectively. It is assumed that $U = V + 2J$ and $J = J^\prime$.
By using the standard mean-field decoupling
\bea
    \left\langle c_{i\alpha\mu}^\dagger c_{i\beta\nu} \right\rangle = \frac{1}{2}\left(n_{i\alpha} + \mu m_{i\alpha} \right) \delta_{\alpha\beta} \delta_{\mu\nu},
\eea
where $\mu = \pm 1$ for up and down spins, respectively, we derive the mean-field interaction term
\bea
    \Ham_I  = \sum_{\bm{k},i,\alpha,\mu} \left( \epsilon_{i\alpha} - \mu \eta_{i\alpha} \right) c_{i\alpha\mu}^\dagger(\bm{k}) c_{i\alpha\mu}(\bm{k}) + C,
\eea
where
\bea
\epsilon_{i\alpha} & = & \frac{U}{2} n_{i\alpha} + \left(V-\frac{J}{2} \right) \sum_{\beta \neq \alpha} n_{i\beta}, \\
\eta_{i\alpha} & = & \frac{U}{2} m_{i\alpha} + \frac{J}{2} \sum_{\beta
\neq \alpha} m_{i\beta},
\eea
and the constant
\bea
   C  & =  &-\frac{U}{4} \sum_{i,\alpha} \left(n_{i\alpha}^2-m_{i\alpha}^2 \right) - \frac{2V -J}{4} \sum_{i,\alpha \neq \beta} n_{i\alpha} n_{i\beta} \nonumber \\ & & + \, \frac{J}{4} \sum_{i,\alpha \neq \beta} m_{i\alpha} m_{i\beta}.
\eea
The Hamiltonian $\Ham = \Ham_K + \Ham_I$ is solved with mean-field parameters $n_{i\alpha}$ and $m_{i\alpha}$ determined self-consistently.
We choose different initial conditions of $n_{i\alpha}$ and $m_{i\alpha}$ that may yield solutions of different spin configurations to determine the phase diagram presented in this paper. We emphasize that this mean-field treatment is not adequate to address strong correlations, but it should produce qualitatively correct result regarding the ground state properties.

\begin{figure}
  \centering
  \includegraphics[width=8cm]{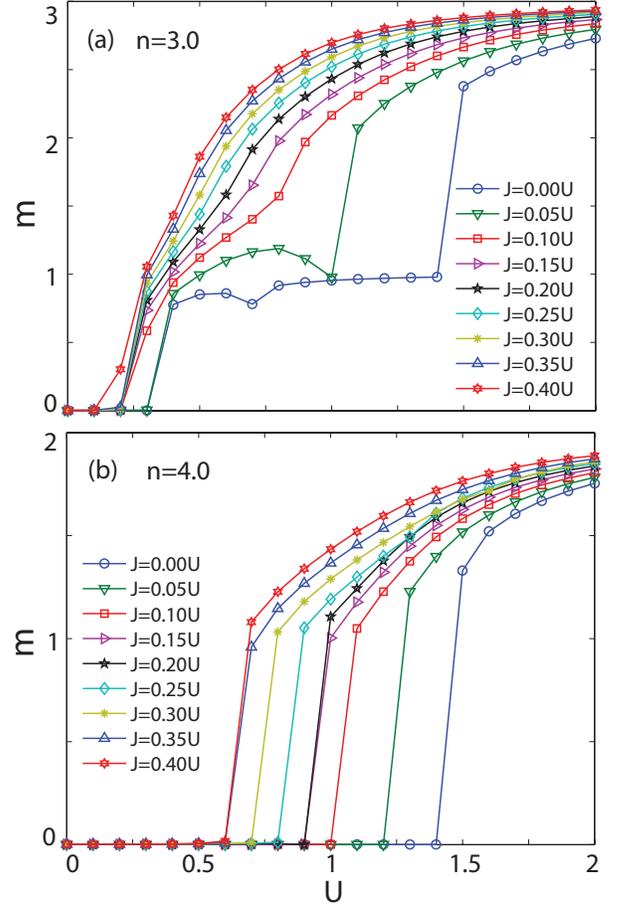}
  \caption{(Color online) The total staggered magnetic moment $m$ of the block antiferromagnetic phase at each Fe site as a function of Coulomb repulsion $U$  for different Hund's couplings $J$. (a) The filling $n = 3.0$; (b) $n = 4.0$.}
  \label{fig:moments}
\end{figure}

We will mainly focus on filling levels between $n=3$ and $n=4$, corresponding to the state of high spin $S=3/2$ and low spin $S=1$, respectively, in the limit of strong interactions.
The motivation for this choice is as follows. The atomic configuration of an Fe ion in FeSe is Fe$^{2+}$, corresponding to six electrons in the five $d$-orbitals.
Due to the crystal field splitting, $e_g$ orbitals have lower energy
than $t_{2g}$ orbitals. Doping K, Tl or Cs into FeSe introduces one
more electron to some of the Fe ions; thus in
(K,Tl,Cs)$_y$Fe$_{2-x}$Se$_2$ there exists a mixture of Fe$^{2+}$ and
Fe$^{+}$. Naively, the large magnetic moment of 3.3$\mu_B$ observed in experiment \cite{bao_w2011}
follows from simply averaging the number of iron ions Fe$^{2+}$ with 4
unpaired electrons
(1 in the $e_g$ and 3 in the $t_{2g}$ orbitals) and Fe$^{+}$ ions with
three (3 in the $t_{2g}$ orbitals). Consequently, we
treat the parent material as having three electrons in the $t_{2g}$ orbitals. Increasing the content of K, Tl or Cs corresponds to electron-doping away from the parent compound.
As a result, the range of filling levels between $n=3$ and $n=4$ in a three-band model is experimentally relevant for this system.

\section{results}
We first plot the total staggered magnetic moment $m = \sum_{\alpha}
m_{i\alpha}$ as a function of the Coulomb repulsion $U$ for different
Hund's exchanges $J$.  Our results are displayed in Fig.~\ref{fig:moments}. For
filling factors of $n=3$, $m$ turns up almost continuously with
increasing $U$, featuring an intermediate regime with metallic block
antiferromagnetic order.  This phase is sandwiched between the
paramagnetic phase at small $U$ and the magnetic insulating phase at
large $U$ [Fig.~\ref{fig:moments}(a)]. The only exception is the small
ratio of $J$ and $U$, which actually signals the presence of competing
phases as we will explain later. However, the situation changes
dramatically for $n=4$. The magnetic moment $m$ turns on abruptly at a
critical value of $U$, where the system undergoes a transition from a
paramagnetic metal to a block antiferromagnetic insulator
[Fig.~\ref{fig:moments}(b)]. Compared with earlier studies
\cite{luo2010} of the same model without any vacancy order, we find
that the intermediate metallic phase with a non-zero magnetic moment
disappears, and the insulating behavior obtains at a much smaller
$U$. Hence, we have confirmed that the presence of vacancy order
indeed enhances electron correlations. Note that the real materials
are possibly located close to the edge of the insulating phase where a
block antiferromagnetic order with a large moment is supported by a relatively small Coulomb repulsion $U$ and an intermediate Hund's coupling $J$.

\begin{figure}
  \centering
  \includegraphics[width=8cm]{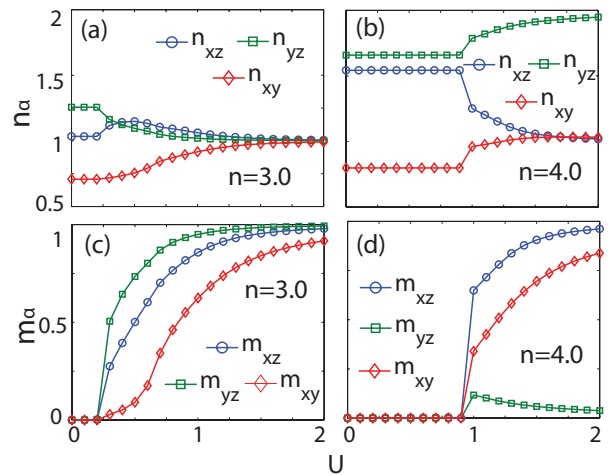}
  \caption{(Color online) (a,b) The occupation number $n_\alpha$ and (c,d) the magnetic moment $m_\alpha$ for each of the three $t_{2g}$ orbitals at site 1 as a function of $U$ for $J=0.20U$. (a,c) $n=3.0$; (b,d) $n=4.0$.}
  \label{fig:charge_spin}
\end{figure}

We now address orbital order. As mentioned earlier, orbital
order of the type shown in Fig.~\ref{fig:schematic} should be present
even in the absence of any interactions.  This is indeed confirmed by
Figs.~\ref{fig:charge_spin}(a) and (b), where the occupation number of
the $d_{yz}$ orbital is larger than that of $d_{xz}$ on site 1 at
$U=0$. Of course, we need to invert the occupation numbers of these
two orbitals on site 2, and so on. We note that the result here is
different from that of Fig.~\ref{fig:schematic} where the $d_{xz}$
orbital is favored on site 1.  This difference is due to the choice of
hopping parameters. Nevertheless, the physical idea remains the same that the vacancy order produces different kinetic energy modulations at the $d_{xz}$ and $d_{yz}$ orbitals, breaking their degeneracy and leading to orbital order. The resultant orbital polarizations, however, will depend on the set of tight-binding hopping parameters chosen.

From Fig.~\ref{fig:charge_spin}, we notice that for $n=3$, the orbital order is reduced once the magnetic order sets in and finally diminishes at large $U$. This novel inverse relation
between orbital polarization ($n_{xz}-n_{yz}$) and magnetic moment can be naturally understood within our model. Since all the three orbitals are singly-occupied with their spins pointing along the same direction in the high-spin $S=3/2$ state, no orbital order can occur and vice versa. On the other hand, for $n=4$, the orbital order is greatly enhanced by the magnetic order.
Because the $d_{xy}$ orbital has a higher band energy than do the $d_{xz}$ and $d_{yz}$ orbitals as shown in the LDA calculations \cite{graser2009},
the system with $n=4$ finally evolves into the state in which $d_{yz}$ is doubly-occupied whereas $d_{xz}$ and $d_{xy}$ are singly-occupied,
consistent with the low-spin $S=1$ state of the largest orbital polarizations.

\begin{figure}
  \centering
    \includegraphics[width=8cm]{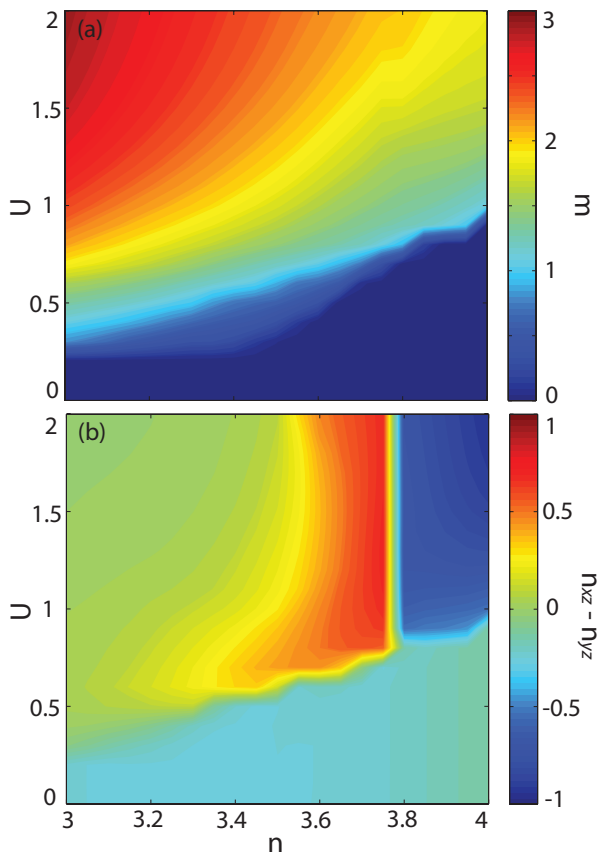}
      \caption{(Color online) (a) The total magnetic moment and (b) the orbital polarization, defined as the occupation number difference between the $d_{xz}$ and $d_{yz}$ orbitals at site 1, as functions of filling $n$ and Coulomb repulsion $U$. We choose the Hund's coupling $J=0.20U$.}
        \label{fig:filling}
	\end{figure}

The discussion above immediately suggests that the phase transition from the high-spin and orbitally-disordered state at filling $n=3$ to the low-spin and
orbitally-ordered state at $n=4$
is non-trivial. In order to simplify our calculations, we fix
$J=0.20U$, consistent with our earlier considerations. From
Fig.~\ref{fig:filling}(a), the total staggered magnetic moment $m$
exhibits a continuous change when we vary the filling level $n$, and
the inverse relation between the orbital polarization and the magnetic
moment remains until $n\approx 3.8$, where a sharp transition of the
orbital polarization occurs for a large enough Coulomb repulsion $U
\gtrsim 0.8$ [Fig.~\ref{fig:filling}(b)]. The system changes from a
$d_{xz}$-polarized state into a state where the $d_{yz}$ orbital
dominates. From the experimental point of view, the two states, although having similar magnetic moments, have opposite orbital polarizations. This observation can help us determine the parameter space of the real material.

\begin{figure}
  \centering
  \includegraphics[width=8cm]{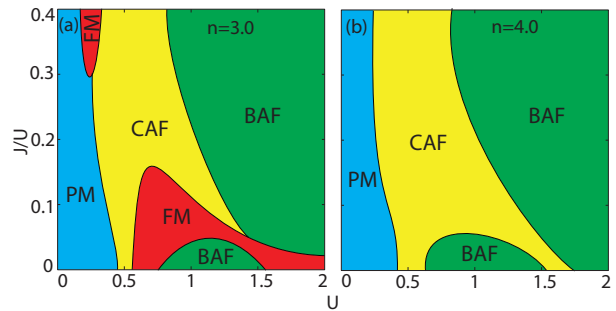}
  \caption{(Color online) The phase diagram obtained by comparing the energies of different mean-field solutions. Refer to the main text for the meanings of the abbreviations. (a) $n=3.0$; (b) $n=4.0$.}
  \label{fig:phase}
\end{figure}

It should also be noted that our study is based on self-consistent mean-field theory. Although the block antiferromagnetic (BAF) phase does emerge as a solution in a large part
of phase diagram, there are still other possible magnetically ordered
states. For this purpose, we consider two other possibilities, the
ferromagnetic (FM) phase where the spins are all aligned in the same
directions, and the checkerboard antiferromagnetic (CAF) phase where
the spins on nearest-neighbor sites are antiparallel. There are
certainly other possible configurations which are ignored here for
simplicity. We find that self-consistent solutions can be obtained for
all the three phases. By comparing the energies of each state, we
obtain the phase diagram shown in Fig.~\ref{fig:phase}, where the
paramagnetic (PM) phase is characterized by a vanishing magnetic
moment. As illustrated by the phase diagram, the block
antiferromagnetic phase nearly always has the lowest energy in the
regime of interest, which confirms that our model does support a
ground state that is consistent with the experimental observations. We
also notice that in the regime where $J/U$ is small, a variety of
phases obtain depending on the parameters, which suggests
the ratio $J/U$ may have an intermediate value in the real materials.

To further demonstrate the stability of the BAF spin configuration, we consider the strong coupling limit, namely a superexchange Heisenberg model due to the presence of the orbital order (see Fig.~\ref{fig:schematic}). For simplicity, the $d_{xy}$ orbital is dropped due to its higher on-site energy \cite{daghofer2010}. We also assume the largest orbital polarization on each site, corresponding to an $S=1/2$ state. The magnetic superexchanges arise from the virtual hopping processes including the nearest-neighbor (NN) $\sigma$-bond $t_1$, $\pi$-bond $t_2$ and the next-nearest-neighbor (NNN) intra-orbital $t_3$, inter-orbital $t_4$, following the definitions of Ref.~\onlinecite{daghofer2010}. We will assume that $t_1>t_2$, which produces the orbital order displayed in Fig.~\ref{fig:schematic}. Straightforward calculations yield
\bea
    J_1 & = &-2 \left(t_1^2 + t_2^2\right) \left( \frac{1}{V-J} - \frac{1}{V+J} \right), \label{eq:j1}\\
    J_1^\prime & =&  \frac{4t_1^2}{U}, \\
    J_2 & = & \frac{4t_3^2}{U} - 4t_4^2\left( \frac{1}{V-J} - \frac{1}{V+J} \right), \\
    J_2^\prime & = & \frac{4t_4^2}{U} - 4t_3^2\left( \frac{1}{V-J} - \frac{1}{V+J} \right),
\eea
where $J_1$, $J_2$, $J_1^\prime$, and $J_2^\prime$ represent the
intra- and inter-block NN and NNN superexchanges, respectively, as
illustrated in Fig.~\ref{fig:schematic}. 

Let's discuss $J_1$ and $J_1^\prime$ first.
It is not surprising that $J_1^\prime$ is always positive because it is related to a superexchange process between two sites occupied by the same orbital.
Interestingly, we find that $J_1$ is always negative due to the Hund's coupling, which can be understood as follows.
Because $J_1$ is the superexchange involving two sites with different orbitals, the intermediate high-energy states can be either spin parallel or spin antiparallel.
If there is no Hund's coupling, these two intermediate states are degenerate and their contributions to the exchange constant cancel each other. This can be checked by setting $J=0$ in Eq.~(\ref{eq:j1}).
Turning on Hund's coupling $J$ favors the spin parallel intermediate state, leading to $J_1 <0$.
Our results of $J_1$ and $J_1^\prime$ can also be understood as the consequence of a generalized Goodenough-Kanamori rule.

The signs of $J_2$ and $J_2^\prime$, however, depend
on the hopping parameters and interaction strengths. But $J_2$ and
$J_2^\prime$ are usually antiferromagnetic because the first term is
proportional to $t^2/U$, thereby winning out over the second term that
scales as $t^2J/V^2$. Compared to earlier LDA results \cite{cao2011}, in which a large antiferromagnetic inter-block NNN exchange $J_2^\prime$ dictates the magnetic ground state, the BAF spin configuration is mostly stabilized by a ferromagnetic intra-block NN $J_1$ and an antiferromagnetic inter-block NN $J_1^\prime$ in our model. Actually the signs of the exchange constants we predicted here agree with the fitting results of recent inelastic neutron scattering experiments\cite{wang2011}, which lends further support to our model.

\section{Summary}
In conclusion, we have studied the phase diagram of the orbital and magnetic orderings
using a three-orbital ($t_{2g}$) tight-binding model with generalized Hubbard interactions
for the recently discovered high-temperature superconductor (K,Tl,Cs)$_y$Fe$_{2-x}$Se$_2$.
The $\sqrt{5}\times\sqrt{5}$ ordering of Fe vacancies has been put
into the calculations explicitly.
We have shown that while the vacancy ordering breaks local
C$_4$ symmetry on each Fe site, thereby yielding an orbitally ordered
state, which has been hinted at by a recent NMR study \cite{torchetti2011}, it also enhances
the correlation effects resulting in magnetic ordering with a large moment.
These two trends compete, leading to a novel inverse relation between
orbital polarization and magnetic moment in the ground state.

We have also derived an effective Heisenberg model with both vacancy and orbital orders in the strong coupling limit.
By superexchange mechanism, we have found that the intra- and inter-block nearest-neighbor ($J_1$, $J_1^\prime$) and
next-nearest-neighbor ($J_2$, $J_2^\prime$) superexchanges do fall into the parameter region which favors the block antiferromagnetic phase as
the ground state, and signs of these exchange constants agree with a recent inelastic neutron scattering experiment.\cite{wang2011}
This provides another strong support for our theory.
Furthermore, we predict that a transition from high-spin to low-spin states together with a
crossover from orbitally-disordered to orbitally-ordered states
will be driven by doping the parent compound with electrons, which might be
verified by further neutron scattering and soft X-ray measurements.

\begin{acknowledgments}
We would like to thank Seungmin Hong and Wei Ku for helpful discussions. This work is supported by NSF DMR-0940992 and the
Center for Emergent Superconductivity, a DOE Energy Frontier Research Center, Grant No.~DE-AC0298CH1088.
\end{acknowledgments}


\begin{thebibliography}{22}%
\makeatletter
\providecommand \@ifxundefined [1]{%
 \@ifx{#1\undefined}
}%
\providecommand \@ifnum [1]{%
 \ifnum #1\expandafter \@firstoftwo
 \else \expandafter \@secondoftwo
 \fi
}%
\providecommand \@ifx [1]{%
 \ifx #1\expandafter \@firstoftwo
 \else \expandafter \@secondoftwo
 \fi
}%
\providecommand \natexlab [1]{#1}%
\providecommand \enquote  [1]{``#1''}%
\providecommand \bibnamefont  [1]{#1}%
\providecommand \bibfnamefont [1]{#1}%
\providecommand \citenamefont [1]{#1}%
\providecommand \href@noop [0]{\@secondoftwo}%
\providecommand \href [0]{\begingroup \@sanitize@url \@href}%
\providecommand \@href[1]{\@@startlink{#1}\@@href}%
\providecommand \@@href[1]{\endgroup#1\@@endlink}%
\providecommand \@sanitize@url [0]{\catcode `\\12\catcode `\$12\catcode
  `\&12\catcode `\#12\catcode `\^12\catcode `\_12\catcode `\%12\relax}%
\providecommand \@@startlink[1]{}%
\providecommand \@@endlink[0]{}%
\providecommand \url  [0]{\begingroup\@sanitize@url \@url }%
\providecommand \@url [1]{\endgroup\@href {#1}{\urlprefix }}%
\providecommand \urlprefix  [0]{URL }%
\providecommand \Eprint [0]{\href }%
\providecommand \doibase [0]{http://dx.doi.org/}%
\providecommand \selectlanguage [0]{\@gobble}%
\providecommand \bibinfo  [0]{\@secondoftwo}%
\providecommand \bibfield  [0]{\@secondoftwo}%
\providecommand \translation [1]{[#1]}%
\providecommand \BibitemOpen [0]{}%
\providecommand \bibitemStop [0]{}%
\providecommand \bibitemNoStop [0]{.\EOS\space}%
\providecommand \EOS [0]{\spacefactor3000\relax}%
\providecommand \BibitemShut  [1]{\csname bibitem#1\endcsname}%
\let\auto@bib@innerbib\@empty
\bibitem [{\citenamefont {Hsu}\ \emph {et~al.}(2008)\citenamefont {Hsu},
  \citenamefont {Luo}, \citenamefont {Yeh}, \citenamefont {Chen}, \citenamefont
  {Huang}, \citenamefont {Wu}, \citenamefont {Lee}, \citenamefont {Huang},
  \citenamefont {Chu}, \citenamefont {Yan},\ and\ \citenamefont
  {Wu}}]{hsu_fc2008}%
  \BibitemOpen
  \bibfield  {author} {\bibinfo {author} {\bibfnamefont {F.-C.}\ \bibnamefont
  {Hsu}}, \bibinfo {author} {\bibfnamefont {J.-Y.}\ \bibnamefont {Luo}},
  \bibinfo {author} {\bibfnamefont {K.-W.}\ \bibnamefont {Yeh}}, \bibinfo
  {author} {\bibfnamefont {T.-K.}\ \bibnamefont {Chen}}, \bibinfo {author}
  {\bibfnamefont {T.-W.}\ \bibnamefont {Huang}}, \bibinfo {author}
  {\bibfnamefont {P.~M.}\ \bibnamefont {Wu}}, \bibinfo {author} {\bibfnamefont
  {Y.-C.}\ \bibnamefont {Lee}}, \bibinfo {author} {\bibfnamefont {Y.-L.}\
  \bibnamefont {Huang}}, \bibinfo {author} {\bibfnamefont {Y.-Y.}\ \bibnamefont
  {Chu}}, \bibinfo {author} {\bibfnamefont {D.-C.}\ \bibnamefont {Yan}}, \ and\
  \bibinfo {author} {\bibfnamefont {M.-K.}\ \bibnamefont {Wu}},\ }\href@noop {}
  {\bibfield  {journal} {\bibinfo  {journal} {PNAS}\ }\textbf {\bibinfo
  {volume} {105}},\ \bibinfo {pages} {14262} (\bibinfo {year}
  {2008})}\BibitemShut {NoStop}%
\bibitem [{\citenamefont {Guo}\ \emph {et~al.}(2010)\citenamefont {Guo},
  \citenamefont {Jin}, \citenamefont {Wang}, \citenamefont {Wang},
  \citenamefont {Zhu}, \citenamefont {Zhou}, \citenamefont {He},\ and\
  \citenamefont {Chen}}]{guo_jg2010}%
  \BibitemOpen
  \bibfield  {author} {\bibinfo {author} {\bibfnamefont {J.}~\bibnamefont
  {Guo}}, \bibinfo {author} {\bibfnamefont {S.}~\bibnamefont {Jin}}, \bibinfo
  {author} {\bibfnamefont {G.}~\bibnamefont {Wang}}, \bibinfo {author}
  {\bibfnamefont {S.}~\bibnamefont {Wang}}, \bibinfo {author} {\bibfnamefont
  {K.}~\bibnamefont {Zhu}}, \bibinfo {author} {\bibfnamefont {T.}~\bibnamefont
  {Zhou}}, \bibinfo {author} {\bibfnamefont {M.}~\bibnamefont {He}}, \ and\
  \bibinfo {author} {\bibfnamefont {X.}~\bibnamefont {Chen}},\ }\href@noop {}
  {\bibfield  {journal} {\bibinfo  {journal} {Phys. Rev. B}\ }\textbf {\bibinfo
  {volume} {82}},\ \bibinfo {pages} {180520} (\bibinfo {year}
  {2010})}\BibitemShut {NoStop}%
\bibitem [{\citenamefont {Mizuguchi}\ \emph {et~al.}(2011)\citenamefont
  {Mizuguchi}, \citenamefont {Takeya}, \citenamefont {Kawasaki}, \citenamefont
  {Ozaki}, \citenamefont {Tsuda}, \citenamefont {Yamaguchi},\ and\
  \citenamefont {Takano}}]{mizuguchi2011}%
  \BibitemOpen
  \bibfield  {author} {\bibinfo {author} {\bibfnamefont {Y.}~\bibnamefont
  {Mizuguchi}}, \bibinfo {author} {\bibfnamefont {H.}~\bibnamefont {Takeya}},
  \bibinfo {author} {\bibfnamefont {Y.}~\bibnamefont {Kawasaki}}, \bibinfo
  {author} {\bibfnamefont {T.}~\bibnamefont {Ozaki}}, \bibinfo {author}
  {\bibfnamefont {S.}~\bibnamefont {Tsuda}}, \bibinfo {author} {\bibfnamefont
  {T.}~\bibnamefont {Yamaguchi}}, \ and\ \bibinfo {author} {\bibfnamefont
  {Y.}~\bibnamefont {Takano}},\ }\href@noop {} {\bibfield  {journal} {\bibinfo
  {journal} {App. Phys. Lett.}\ }\textbf {\bibinfo {volume} {98}},\ \bibinfo
  {pages} {042511} (\bibinfo {year} {2011})}\BibitemShut {NoStop}%
\bibitem [{\citenamefont {Wang}\ \emph {et~al.}(2011)\citenamefont {Wang},
  \citenamefont {Ying}, \citenamefont {Yan}, \citenamefont {Liu}, \citenamefont
  {Luo}, \citenamefont {Li}, \citenamefont {Wang}, \citenamefont {Zhang},
  \citenamefont {Ye}, \citenamefont {Cheng}, \citenamefont {Xiang},\ and\
  \citenamefont {Chen}}]{wang_af2011}%
  \BibitemOpen
  \bibfield  {author} {\bibinfo {author} {\bibfnamefont {A.~F.}\ \bibnamefont
  {Wang}}, \bibinfo {author} {\bibfnamefont {J.~J.}\ \bibnamefont {Ying}},
  \bibinfo {author} {\bibfnamefont {Y.~J.}\ \bibnamefont {Yan}}, \bibinfo
  {author} {\bibfnamefont {R.~H.}\ \bibnamefont {Liu}}, \bibinfo {author}
  {\bibfnamefont {X.~G.}\ \bibnamefont {Luo}}, \bibinfo {author} {\bibfnamefont
  {Z.~Y.}\ \bibnamefont {Li}}, \bibinfo {author} {\bibfnamefont {X.~F.}\
  \bibnamefont {Wang}}, \bibinfo {author} {\bibfnamefont {M.}~\bibnamefont
  {Zhang}}, \bibinfo {author} {\bibfnamefont {G.~J.}\ \bibnamefont {Ye}},
  \bibinfo {author} {\bibfnamefont {P.}~\bibnamefont {Cheng}}, \bibinfo
  {author} {\bibfnamefont {Z.~J.}\ \bibnamefont {Xiang}}, \ and\ \bibinfo
  {author} {\bibfnamefont {X.~H.}\ \bibnamefont {Chen}},\ }\href@noop {}
  {\bibfield  {journal} {\bibinfo  {journal} {Phys. Rev. B}\ }\textbf {\bibinfo
  {volume} {83}},\ \bibinfo {pages} {060512} (\bibinfo {year}
  {2011})}\BibitemShut {NoStop}%
\bibitem [{\citenamefont {Fang}\ \emph {et~al.}(2011)\citenamefont {Fang},
  \citenamefont {Wang}, \citenamefont {Dong}, \citenamefont {Li}, \citenamefont
  {Feng}, \citenamefont {Chen},\ and\ \citenamefont {Yuan}}]{fang_mh2011}%
  \BibitemOpen
  \bibfield  {author} {\bibinfo {author} {\bibfnamefont {M.-H.}\ \bibnamefont
  {Fang}}, \bibinfo {author} {\bibfnamefont {H.-D.}\ \bibnamefont {Wang}},
  \bibinfo {author} {\bibfnamefont {C.-H.}\ \bibnamefont {Dong}}, \bibinfo
  {author} {\bibfnamefont {Z.-J.}\ \bibnamefont {Li}}, \bibinfo {author}
  {\bibfnamefont {C.-M.}\ \bibnamefont {Feng}}, \bibinfo {author}
  {\bibfnamefont {J.}~\bibnamefont {Chen}}, \ and\ \bibinfo {author}
  {\bibfnamefont {H.~Q.}\ \bibnamefont {Yuan}},\ }\href@noop {} {\bibfield
  {journal} {\bibinfo  {journal} {EPL}\ }\textbf {\bibinfo {volume} {94}},\
  \bibinfo {pages} {27009} (\bibinfo {year} {2011})}\BibitemShut {NoStop}%
\bibitem [{\citenamefont {Bacsa}\ \emph {et~al.}(2011)\citenamefont {Bacsa},
  \citenamefont {Ganin}, \citenamefont {Takabayashi}, \citenamefont
  {Christensen}, \citenamefont {Prassides}, \citenamefont {Rosseinsky},\ and\
  \citenamefont {Claridge}}]{bacsa2011}%
  \BibitemOpen
  \bibfield  {author} {\bibinfo {author} {\bibfnamefont {J.}~\bibnamefont
  {Bacsa}}, \bibinfo {author} {\bibfnamefont {A.~Y.}\ \bibnamefont {Ganin}},
  \bibinfo {author} {\bibfnamefont {Y.}~\bibnamefont {Takabayashi}}, \bibinfo
  {author} {\bibfnamefont {K.~E.}\ \bibnamefont {Christensen}}, \bibinfo
  {author} {\bibfnamefont {K.}~\bibnamefont {Prassides}}, \bibinfo {author}
  {\bibfnamefont {M.~J.}\ \bibnamefont {Rosseinsky}}, \ and\ \bibinfo {author}
  {\bibfnamefont {J.~B.}\ \bibnamefont {Claridge}},\ }\href@noop {} {\bibfield
  {journal} {\bibinfo  {journal} {Chem. Sci.}\ }\textbf {\bibinfo {volume}
  {2}},\ \bibinfo {pages} {1054} (\bibinfo {year} {2011})}\BibitemShut
  {NoStop}%
\bibitem [{\citenamefont {Zavalij}\ \emph {et~al.}(2011)\citenamefont
  {Zavalij}, \citenamefont {Bao}, \citenamefont {Wang}, \citenamefont {Ying},
  \citenamefont {Chen}, \citenamefont {Wang}, \citenamefont {He}, \citenamefont
  {Wang}, \citenamefont {Chen}, \citenamefont {Hsieh}, \citenamefont {Huang},\
  and\ \citenamefont {Green}}]{zavalij2011}%
  \BibitemOpen
  \bibfield  {author} {\bibinfo {author} {\bibfnamefont {P.}~\bibnamefont
  {Zavalij}}, \bibinfo {author} {\bibfnamefont {W.}~\bibnamefont {Bao}},
  \bibinfo {author} {\bibfnamefont {X.~F.}\ \bibnamefont {Wang}}, \bibinfo
  {author} {\bibfnamefont {J.~J.}\ \bibnamefont {Ying}}, \bibinfo {author}
  {\bibfnamefont {X.~H.}\ \bibnamefont {Chen}}, \bibinfo {author}
  {\bibfnamefont {D.~M.}\ \bibnamefont {Wang}}, \bibinfo {author}
  {\bibfnamefont {J.~B.}\ \bibnamefont {He}}, \bibinfo {author} {\bibfnamefont
  {X.~Q.}\ \bibnamefont {Wang}}, \bibinfo {author} {\bibfnamefont {G.~F.}\
  \bibnamefont {Chen}}, \bibinfo {author} {\bibfnamefont {P.-Y.}\ \bibnamefont
  {Hsieh}}, \bibinfo {author} {\bibfnamefont {Q.}~\bibnamefont {Huang}}, \ and\
  \bibinfo {author} {\bibfnamefont {M.~A.}\ \bibnamefont {Green}},\ }\href@noop
  {} {\bibfield  {journal} {\bibinfo  {journal} {Phys. Rev. B}\ }\textbf
  {\bibinfo {volume} {83}},\ \bibinfo {pages} {132509} (\bibinfo {year}
  {2011})}\BibitemShut {NoStop}%
\bibitem [{\citenamefont {Bao}\ \emph {et~al.}(2011)\citenamefont {Bao},
  \citenamefont {Huang}, \citenamefont {Chen}, \citenamefont {Green},
  \citenamefont {Wang}, \citenamefont {He}, \citenamefont {Wang},\ and\
  \citenamefont {Qiu}}]{bao_w2011}%
  \BibitemOpen
  \bibfield  {author} {\bibinfo {author} {\bibfnamefont {W.}~\bibnamefont
  {Bao}}, \bibinfo {author} {\bibfnamefont {Q.}~\bibnamefont {Huang}}, \bibinfo
  {author} {\bibfnamefont {G.~F.}\ \bibnamefont {Chen}}, \bibinfo {author}
  {\bibfnamefont {M.~A.}\ \bibnamefont {Green}}, \bibinfo {author}
  {\bibfnamefont {D.~M.}\ \bibnamefont {Wang}}, \bibinfo {author}
  {\bibfnamefont {J.~B.}\ \bibnamefont {He}}, \bibinfo {author} {\bibfnamefont
  {X.~Q.}\ \bibnamefont {Wang}}, \ and\ \bibinfo {author} {\bibfnamefont
  {Y.}~\bibnamefont {Qiu}},\ }\href@noop {} {\bibfield  {journal} {\bibinfo
  {journal} {Chin. Phys. Lett.}\ }\textbf {\bibinfo {volume} {28}},\ \bibinfo
  {pages} {086104} (\bibinfo {year} {2011})}\BibitemShut {NoStop}%
\bibitem [{\citenamefont {Lee}\ and\ \citenamefont {Phillips}(2011)}]{wclee}%
  \BibitemOpen
  \bibfield  {author} {\bibinfo {author} {\bibfnamefont {W.-C.}\ \bibnamefont
  {Lee}}\ and\ \bibinfo {author} {\bibfnamefont {P.~W.}\ \bibnamefont
  {Phillips}},\ }\href@noop {} {\bibfield  {journal} {\bibinfo  {journal}
  {Phys. Rev. B}\ }\textbf {\bibinfo {volume} {84}},\ \bibinfo {pages} {115101}
  (\bibinfo {year} {2011})}\BibitemShut {NoStop}%
\bibitem [{\citenamefont {Zhou}\ \emph {et~al.}()\citenamefont {Zhou},
  \citenamefont {Chen}, \citenamefont {Guo}, \citenamefont {Wang},
  \citenamefont {Lai}, \citenamefont {Wang}, \citenamefont {Jin},\ and\
  \citenamefont {Zhu}}]{zhou_tt2011}%
  \BibitemOpen
  \bibfield  {author} {\bibinfo {author} {\bibfnamefont {T.}~\bibnamefont
  {Zhou}}, \bibinfo {author} {\bibfnamefont {X.}~\bibnamefont {Chen}}, \bibinfo
  {author} {\bibfnamefont {J.}~\bibnamefont {Guo}}, \bibinfo {author}
  {\bibfnamefont {G.}~\bibnamefont {Wang}}, \bibinfo {author} {\bibfnamefont
  {X.}~\bibnamefont {Lai}}, \bibinfo {author} {\bibfnamefont {S.}~\bibnamefont
  {Wang}}, \bibinfo {author} {\bibfnamefont {S.}~\bibnamefont {Jin}}, \ and\
  \bibinfo {author} {\bibfnamefont {K.}~\bibnamefont {Zhu}},\ }\href@noop {} {\
  }\Eprint {http://arxiv.org/abs/arXiv:1102.3506} {arXiv:1102.3506}
  \BibitemShut {NoStop}%
\bibitem [{\citenamefont {Yu}\ \emph {et~al.}(2011)\citenamefont {Yu},
  \citenamefont {Zhu},\ and\ \citenamefont {Si}}]{yu_r2011}%
  \BibitemOpen
  \bibfield  {author} {\bibinfo {author} {\bibfnamefont {R.}~\bibnamefont
  {Yu}}, \bibinfo {author} {\bibfnamefont {J.-X.}\ \bibnamefont {Zhu}}, \ and\
  \bibinfo {author} {\bibfnamefont {Q.}~\bibnamefont {Si}},\ }\href@noop {}
  {\bibfield  {journal} {\bibinfo  {journal} {Phys. Rev. Lett.}\ }\textbf
  {\bibinfo {volume} {106}},\ \bibinfo {pages} {186401} (\bibinfo {year}
  {2011})}\BibitemShut {NoStop}%
\bibitem [{\citenamefont {Zhou}\ \emph {et~al.}(2011)\citenamefont {Zhou},
  \citenamefont {Xu}, \citenamefont {Zhang},\ and\ \citenamefont
  {Chen}}]{zhou_y2011}%
  \BibitemOpen
  \bibfield  {author} {\bibinfo {author} {\bibfnamefont {Y.}~\bibnamefont
  {Zhou}}, \bibinfo {author} {\bibfnamefont {D.-H.}\ \bibnamefont {Xu}},
  \bibinfo {author} {\bibfnamefont {F.-C.}\ \bibnamefont {Zhang}}, \ and\
  \bibinfo {author} {\bibfnamefont {W.-Q.}\ \bibnamefont {Chen}},\ }\href@noop
  {} {\bibfield  {journal} {\bibinfo  {journal} {EPL}\ }\textbf {\bibinfo
  {volume} {95}},\ \bibinfo {pages} {17003} (\bibinfo {year}
  {2011})}\BibitemShut {NoStop}%
\bibitem [{\citenamefont {Cao}\ and\ \citenamefont {Dai}(2011)}]{cao2011}%
  \BibitemOpen
  \bibfield  {author} {\bibinfo {author} {\bibfnamefont {C.}~\bibnamefont
  {Cao}}\ and\ \bibinfo {author} {\bibfnamefont {J.}~\bibnamefont {Dai}},\
  }\href@noop {} {\bibfield  {journal} {\bibinfo  {journal} {Phys. Rev. Lett.}\
  }\textbf {\bibinfo {volume} {107}},\ \bibinfo {pages} {056401} (\bibinfo
  {year} {2011})}\BibitemShut {NoStop}%
\bibitem [{\citenamefont {Lu}\ and\ \citenamefont {Dai}()}]{lu_f2011}%
  \BibitemOpen
  \bibfield  {author} {\bibinfo {author} {\bibfnamefont {F.}~\bibnamefont
  {Lu}}\ and\ \bibinfo {author} {\bibfnamefont {X.}~\bibnamefont {Dai}},\
  }\href@noop {} {\ }\Eprint {http://arxiv.org/abs/arXiv:1103.5521}
  {arXiv:1103.5521} \BibitemShut {NoStop}%
\bibitem [{\citenamefont {Yu}\ \emph {et~al.}()\citenamefont {Yu},
  \citenamefont {Goswami},\ and\ \citenamefont {Si}}]{yu_r2011b}%
  \BibitemOpen
  \bibfield  {author} {\bibinfo {author} {\bibfnamefont {R.}~\bibnamefont
  {Yu}}, \bibinfo {author} {\bibfnamefont {P.}~\bibnamefont {Goswami}}, \ and\
  \bibinfo {author} {\bibfnamefont {Q.}~\bibnamefont {Si}},\ }\href@noop {} {\
  }\Eprint {http://arxiv.org/abs/arXiv:1104.1445} {arXiv:1104.1445}
  \BibitemShut {NoStop}%
\bibitem [{\citenamefont {Xu}\ \emph {et~al.}()\citenamefont {Xu},
  \citenamefont {Fang}, \citenamefont {Liu},\ and\ \citenamefont
  {Hu}}]{xu_b2011}%
  \BibitemOpen
  \bibfield  {author} {\bibinfo {author} {\bibfnamefont {B.}~\bibnamefont
  {Xu}}, \bibinfo {author} {\bibfnamefont {C.}~\bibnamefont {Fang}}, \bibinfo
  {author} {\bibfnamefont {W.}~\bibnamefont {Liu}}, \ and\ \bibinfo {author}
  {\bibfnamefont {J.}~\bibnamefont {Hu}},\ }\href@noop {} {\ }\Eprint
  {http://arxiv.org/abs/arXiv:1104.1848} {arXiv:1104.1848} \BibitemShut
  {NoStop}%
\bibitem [{\citenamefont {Chen}\ \emph {et~al.}(2011)\citenamefont {Chen},
  \citenamefont {Cao},\ and\ \citenamefont {Dai}}]{chen2011}%
  \BibitemOpen
  \bibfield  {author} {\bibinfo {author} {\bibfnamefont {H.}~\bibnamefont
  {Chen}}, \bibinfo {author} {\bibfnamefont {C.}~\bibnamefont {Cao}}, \ and\
  \bibinfo {author} {\bibfnamefont {J.}~\bibnamefont {Dai}},\ }\href@noop {}
  {\bibfield  {journal} {\bibinfo  {journal} {Phys. Rev. B}\ }\textbf {\bibinfo
  {volume} {83}},\ \bibinfo {pages} {180413} (\bibinfo {year}
  {2011})}\BibitemShut {NoStop}%
\bibitem [{\citenamefont {Torchetti}\ \emph {et~al.}(2011)\citenamefont
  {Torchetti}, \citenamefont {Fu}, \citenamefont {Christensen}, \citenamefont
  {Nelson}, \citenamefont {Imai}, \citenamefont {Lei},\ and\ \citenamefont
  {Petrovic}}]{torchetti2011}%
  \BibitemOpen
  \bibfield  {author} {\bibinfo {author} {\bibfnamefont {D.~A.}\ \bibnamefont
  {Torchetti}}, \bibinfo {author} {\bibfnamefont {M.}~\bibnamefont {Fu}},
  \bibinfo {author} {\bibfnamefont {D.~C.}\ \bibnamefont {Christensen}},
  \bibinfo {author} {\bibfnamefont {K.~J.}\ \bibnamefont {Nelson}}, \bibinfo
  {author} {\bibfnamefont {T.}~\bibnamefont {Imai}}, \bibinfo {author}
  {\bibfnamefont {H.~C.}\ \bibnamefont {Lei}}, \ and\ \bibinfo {author}
  {\bibfnamefont {C.}~\bibnamefont {Petrovic}},\ }\href@noop {} {\bibfield
  {journal} {\bibinfo  {journal} {Phys. Rev. B}\ }\textbf {\bibinfo {volume}
  {83}},\ \bibinfo {pages} {104508} (\bibinfo {year} {2011})}\BibitemShut
  {NoStop}%
\bibitem [{\citenamefont {Daghofer}\ \emph {et~al.}(2010)\citenamefont
  {Daghofer}, \citenamefont {Nicholson}, \citenamefont {Moreo},\ and\
  \citenamefont {Dagotto}}]{daghofer2010}%
  \BibitemOpen
  \bibfield  {author} {\bibinfo {author} {\bibfnamefont {M.}~\bibnamefont
  {Daghofer}}, \bibinfo {author} {\bibfnamefont {A.}~\bibnamefont {Nicholson}},
  \bibinfo {author} {\bibfnamefont {A.}~\bibnamefont {Moreo}}, \ and\ \bibinfo
  {author} {\bibfnamefont {E.}~\bibnamefont {Dagotto}},\ }\href@noop {}
  {\bibfield  {journal} {\bibinfo  {journal} {Phys. Rev. B}\ }\textbf {\bibinfo
  {volume} {81}},\ \bibinfo {pages} {014511} (\bibinfo {year}
  {2010})}\BibitemShut {NoStop}%
\bibitem [{\citenamefont {Luo}\ \emph {et~al.}(2010)\citenamefont {Luo},
  \citenamefont {Martins}, \citenamefont {Yao}, \citenamefont {Daghofer},
  \citenamefont {Yu}, \citenamefont {Moreo},\ and\ \citenamefont
  {Dagotto}}]{luo2010}%
  \BibitemOpen
  \bibfield  {author} {\bibinfo {author} {\bibfnamefont {Q.}~\bibnamefont
  {Luo}}, \bibinfo {author} {\bibfnamefont {G.}~\bibnamefont {Martins}},
  \bibinfo {author} {\bibfnamefont {D.-X.}\ \bibnamefont {Yao}}, \bibinfo
  {author} {\bibfnamefont {M.}~\bibnamefont {Daghofer}}, \bibinfo {author}
  {\bibfnamefont {R.}~\bibnamefont {Yu}}, \bibinfo {author} {\bibfnamefont
  {A.}~\bibnamefont {Moreo}}, \ and\ \bibinfo {author} {\bibfnamefont
  {E.}~\bibnamefont {Dagotto}},\ }\href@noop {} {\bibfield  {journal} {\bibinfo
   {journal} {Phys. Rev. B}\ }\textbf {\bibinfo {volume} {82}},\ \bibinfo
  {pages} {104508} (\bibinfo {year} {2010})}\BibitemShut {NoStop}%
\bibitem [{\citenamefont {Graser}\ \emph {et~al.}(2009)\citenamefont {Graser},
  \citenamefont {Maier}, \citenamefont {Hirschfeld},\ and\ \citenamefont
  {Scalapino}}]{graser2009}%
  \BibitemOpen
  \bibfield  {author} {\bibinfo {author} {\bibfnamefont {S.}~\bibnamefont
  {Graser}}, \bibinfo {author} {\bibfnamefont {T.~A.}\ \bibnamefont {Maier}},
  \bibinfo {author} {\bibfnamefont {P.~J.}\ \bibnamefont {Hirschfeld}}, \ and\
  \bibinfo {author} {\bibfnamefont {D.~J.}\ \bibnamefont {Scalapino}},\
  }\href@noop {} {\bibfield  {journal} {\bibinfo  {journal} {New J. Phys.}\
  }\textbf {\bibinfo {volume} {11}},\ \bibinfo {pages} {025016} (\bibinfo
  {year} {2009})}\BibitemShut {NoStop}%
\bibitem [{\citenamefont {Wang}\ \emph {et~al.}()\citenamefont {Wang},
  \citenamefont {Fang}, \citenamefont {Yao}, \citenamefont {Tan}, \citenamefont
  {Harriger}, \citenamefont {Song}, \citenamefont {Netherton}, \citenamefont
  {Zhang}, \citenamefont {Wang}, \citenamefont {Stone}, \citenamefont {Tian},
  \citenamefont {Hu},\ and\ \citenamefont {Dai}}]{wang2011}%
  \BibitemOpen
  \bibfield  {author} {\bibinfo {author} {\bibfnamefont {M.}~\bibnamefont
  {Wang}}, \bibinfo {author} {\bibfnamefont {C.}~\bibnamefont {Fang}}, \bibinfo
  {author} {\bibfnamefont {D.-X.}\ \bibnamefont {Yao}}, \bibinfo {author}
  {\bibfnamefont {G.}~\bibnamefont {Tan}}, \bibinfo {author} {\bibfnamefont
  {L.~W.}\ \bibnamefont {Harriger}}, \bibinfo {author} {\bibfnamefont
  {Y.}~\bibnamefont {Song}}, \bibinfo {author} {\bibfnamefont {T.}~\bibnamefont
  {Netherton}}, \bibinfo {author} {\bibfnamefont {C.}~\bibnamefont {Zhang}},
  \bibinfo {author} {\bibfnamefont {M.}~\bibnamefont {Wang}}, \bibinfo {author}
  {\bibfnamefont {M.~B.}\ \bibnamefont {Stone}}, \bibinfo {author}
  {\bibfnamefont {W.}~\bibnamefont {Tian}}, \bibinfo {author} {\bibfnamefont
  {J.}~\bibnamefont {Hu}}, \ and\ \bibinfo {author} {\bibfnamefont
  {P.}~\bibnamefont {Dai}},\ }\href@noop {} {\ }\Eprint
  {http://arxiv.org/abs/arXiv:1105.4675} {arXiv:1105.4675} \BibitemShut
  {NoStop}%
\end{thebibliography}

%

\end{document}